\documentclass[preprint,3p]{elsarticle} 
\usepackage{multicol,caption}
\usepackage{natbib}
\usepackage{amssymb}
\usepackage[english]{babel}
\usepackage[intlimits,sumlimits]{amsmath}
\usepackage{graphicx}
\usepackage{color}
\usepackage{hyperref}
\usepackage{lineno}
\usepackage{etoolbox}

\newenvironment{Figure}
  {\par\medskip\noindent\minipage{\linewidth}}
  {\endminipage\par\medskip}

\newcommand{\x}{X(3872)}
\newcommand{\EE}{e^+e^-}
\newcommand{\GEE}{\Gamma_{ee}}
\newcommand{\syspsip}{4.5}
\newcommand{\sysx}{6.1}
\newcommand{\sysErrpsip}{99}
\newcommand{\statErrpsip}{18}
\newcommand{\geepsip}{2213}
\newcommand{\geBrX}{0.13}
\newcommand{\geBrXnoSys}{0.125}
\newcommand{\geX}{4.3}
\newcommand{\improve}{46}
\journal{Physics Letters B}

\begin{document}
\modulolinenumbers[2]

\begin{frontmatter}

\title{An Improved Limit for $\GEE$ of $\x$ and $\GEE$ Measurement of $\psi(3686)$}

\author{
      M.~Ablikim$^{1}$, M.~N.~Achasov$^{9,a}$, X.~C.~Ai$^{1}$,
      O.~Albayrak$^{5}$, M.~Albrecht$^{4}$, D.~J.~Ambrose$^{44}$,
      A.~Amoroso$^{48A,48C}$, F.~F.~An$^{1}$, Q.~An$^{45}$,
      J.~Z.~Bai$^{1}$
% ...
      \\
      \vspace{0.2cm}
      (BESIII Collaboration)\\
      \vspace{0.2cm} {\it
        $^{1}$ Institute of High Energy Physics, Beijing 100049, People's Republic of China\\
% ...
$^{4}$ Bochum Ruhr-University, D-44780 Bochum, Germany\\
        $^{5}$ Carnegie Mellon University, Pittsburgh, Pennsylvania 15213, USA\\
% ...
$^{9}$ G.I. Budker Institute of Nuclear Physics SB RAS (BINP), Novosibirsk 630090, Russia\\
% ...
$^{44}$ University of Rochester, Rochester, New York 14627, USA\\
        $^{45}$ University of Science and Technology of China, Hefei 230026, People's Republic of China\\
% ...
$^{48}$ (A)University of Turin, I-10125, Turin, Italy; (B)University of Eastern Piedmont, I-15121, Alessandria, Italy; (C)INFN, I-10125, Turin, Italy\\
% ...
$^{a}$ Also at the Novosibirsk State University, Novosibirsk, 630090, Russia\\
% ...
      }
}

\vspace{0.4cm}

\begin{abstract}
Using the data sets taken at center-of-mass energies above 4~GeV by the BESIII detector at the BEPCII storage ring, we search for the reaction $e^+e^-\rightarrow\gamma_{\rm ISR} X(3872)\rightarrow\gamma_{\rm ISR}\pi^+\pi^-J/\psi$ via the Initial State Radiation technique.
The production of a resonance with quantum numbers $J^{PC}=1^{++}$
such as the $X(3872)$ via single photon $\EE$ annihilation is forbidden, but is allowed by a next-to-leading order box diagram.
We do not observe a significant signal of $X(3872)$, and therefore give an upper limit for the electronic width times the branching fraction $\Gamma_{ee}^{\x}\mathcal{B}(\x\to\pi^+\pi^-J/\psi)<\geBrX$~eV at the 90\% confidence level. This measurement improves upon existing limits by a factor of \improve. Using the same final state, we also measure the electronic width of the $\psi(3686)$ to be $\Gamma_{ee}^{\psi(3686)}=\geepsip\pm \statErrpsip_\text{stat}\pm\sysErrpsip_\text{sys}\,\text{eV}$~.
\end{abstract}


\end{frontmatter}

\linenumbers
\begin{multicols}{2}

\section{Introduction}
The $\x$ resonance was observed in 2003 by Belle~\cite{belle} in the decay channel $\pi^+\pi^-J/\psi$.  The existence of this state was later confirmed by several other experiments~\cite{obs1,obs2,obs3,Aaij:2013zoa,bes3x3872}. The observation of the decay channel $X(3872)\rightarrow\gamma J/\psi$ implies that the state has even C-parity~\cite{Aaij:2013zoa,cparity1,cparity2}. The quantum numbers were finally determined to be $J^{PC}=1^{++}$ ~\cite{Aaij:2013zoa,Abulencia:2006ma}.
However, the intrinsic nature of the resonance is still unknown and has led to many conjectures.
It is a good candidate for a tetraquark state but also for a meson molecule as its mass is close to the $D^0\bar{D}^{*0}$ threshold~\cite{th0}. The recent observation of the decay $Y(4260)\rightarrow\gamma \x$ by BESIII~\cite{bes3x3872} implies that the $\x$ could be a meson molecule, as suggested by a model dependent calculation~\cite{molecule}. On the other hand, the large decay rate of \mbox{$\x\to\gamma\psi(3686)$} observed by BaBar and LHCb, compared to $\x\to\gamma J/\psi$ hints at a tetraquark state explanation ~\cite{cparity2,Aubert:2008ae,LHCbtetra}. One of the interesting quantities, which may help to reveal the structure of the $\x$ is its electronic width $\GEE$. A recent order-of-magnitude calculation using a Vector Meson Dominance model predicts $\GEE^{\x}\approx 0.03~\text{eV}$~\cite{Denig:2014fha}, without any prior assumption regarding the nature of the $\x$.
For comparison, calculations for the $\GEE$ of the ordinary $1^{++}$ charmonium state $\chi_{c1}$ have been carried out~\cite{Kuhn:1979bb} and the electronic width is found to be in the range between $0.044\,\text{eV}$ and $0.46\,\text{eV}$. This was also confirmed in a more recent calculation~\cite{Denig:2014fha}.

The current upper limit for $\Gamma_{ee}^{X(3872)}$ is at the $\mathcal{O}(10^2)$~eV level~\cite{PDG},
which is three orders of magnitude larger than the theoretical prediction.
The aim of this work is to obtain a significantly improved experimental value for the electronic width of $X(3872)$ that may be contrasted with predictions of $\GEE$ within various theoretical models making different assumptions regarding the nature of the $\x$.

\begin{Figure}
\centering
\includegraphics[width=0.66\linewidth]{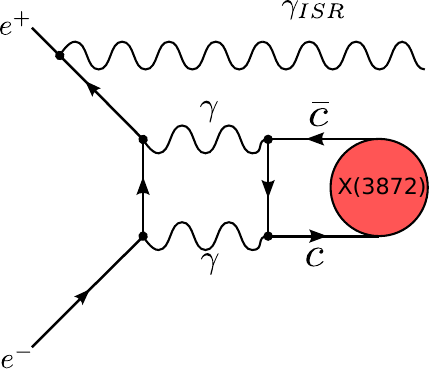}
\captionof{figure}{ISR production of $\x$ via a box diagram.}
\label{box}
\end{Figure}

The production of a $1^{++}$ resonance has never been observed in $\EE$ annihilation so far.
Such a process may occur via a two-photon box diagram as depicted in Fig.~\ref{box}. In order to search for a possible signal we analyze data taken by the BESIII detector at center-of-mass (c.m.) energies above 3.872~GeV, using the Initial State Radiation (ISR) technique.
The ISR photon reduces the available c.m.\ energy, such that the $\x$ can be produced resonantly via the two-photon process.
In the process \mbox{$\EE\to\gamma_{\rm ISR}\x$} we search for the $\x$ in its decay to \mbox{$\pi^{+}\pi^{-}J/\psi$} with \mbox{$J/\psi\to\ell^{+}\ell^{-}$} ($\ell=\mu$ and $e$).
The $\pi^+\pi^-J/\psi$ mass spectrum is expected to be dominated by the well known process \mbox{$\EE\to\gamma_{\rm ISR}\psi(3686)$}.

\section{BESIII Detector, Data and Monte Carlo}
BESIII is a general purpose detector, covering 93\% of the solid angle. It is operating at the $\EE$ double-ring collider BEPCII. A detailed description of the facilities is given in Ref.~\cite{detector}. BESIII consists of four main components: (a) The helium-based 43 layer main drift chamber (MDC) provides an average single-hit resolution of $135\,\mu m$, and a momentum resolution of 0.5\% for charged-particle at 1~GeV/c in a 1~T magnetic field. (b) The electromagnetic calorimeter (EMC) consists of 6240 CsI(Tl) crystals, arrayed in a cylindrical structure (barrel) and two endcaps. The energy resolution for 1.0~GeV photons is 2.5\% (5\%) in the barrel (endcaps), while the position resolution is 6~mm (9~mm) in the barrel (endcaps). (c) The time-of-fight system (TOF) is constructed of 5~cm thick plastic scintillators and includes 88 detectors of 2.4~m length in two layers in the barrel and 96 fan-shaped detectors in the endcaps. The barrel (endcap) time resolution of 80~ps (110~ps) provides 2 sigma $K/\pi$ separation for momenta up to about 1.0~GeV/c. (d) The muon counter (MUC) consists of resistive plate chambers in nine barrel and eight endcap layers. It is incorporated in the return iron of the superconducting magnet. Its position resolution is about 2~cm.

A GEANT4 \cite{mc1,mc2} based detector simulation package is used to model the detector response.
This analysis is based on four data samples taken at c.m.\ energies of 4.009~GeV, 4.230~GeV, 4.260~GeV and 4.360~GeV by the BESIII detector. The integrated luminosity of each data sample is listed in Table~\ref{bwvalues}. The total integrated luminosity is $\mathcal{L}_\text{tot}=2.94\,{\rm fb}^{-1}$.
We simulate the $e^+e^-\to\x\gamma_{\rm ISR}$ signal process using {\sc evtgen}~\cite{Lange:2001uf,Ping:2008zz}, which invokes the {\sc vectorisr} generator model~\cite{VECISR} for the ISR process and the common $\rho J/\psi$ model for the decay \mbox{$\x\to\pi^{+}\pi^{-}J/\psi$}. The Monte Carlo (MC) simulation of the $e^+e^-\to\gamma_{\rm ISR}\psi(3686)$ process was performed using the {\sc phokhara} generator~\cite{Czyz:2010hj}. For the background study we simulate the $e^+e^-\to\eta J/\psi$ process with {\sc evtgen} and the $e^+e^-\to\gamma_{\rm ISR}\pi^+\pi^-\pi^+\pi^-$ process with {\sc phokhara}.

\section{Event Selection}
For the event selection, we require four charged tracks with net charge zero. The point of closest approach to the $\EE$ interaction point is required to be within $\pm 10$~cm in the beam direction and $1$~cm in the plane perpendicular to the beam direction.
As the $J/\psi$ resonance carries most of the total momentum, the final state leptons can be distinguished from pions by their momenta in the lab frame. Tracks with momentum $p>1\,\text{GeV}/c$ in the lab frame are identified as leptons, whereas tracks with $p<600\,\text{MeV}/c$ are identified as pions. The particle identification for leptons is achieved by measuring the ratio of the energy deposited in the EMC divided by the track's momentum measured in the MDC ($E/p$).
If $E/p>0.4$, we assume the lepton to be an electron, otherwise it is considered a muon candidate. The $E/p$ distributions of data and MC agree well, and MC studies show that the background for $J/\psi\rightarrow\EE$ is negligible.
The resolution of the invariant mass of the lepton pairs is $16\,\text{MeV}/c^2$.
We require their invariant mass $M(\ell^{+}\ell^{-})$ to be within \mbox{$3.05\leq M(\ell^{+}\ell^{-})\leq 3.14\,\text{GeV}/c^2$} for the $J/\psi$ signal selection.
Furthermore the opening angle between the two pion tracks is required to satisfy $\cos\alpha_{\pi\pi}\leq 0.6$ to remove background from $\EE\to\eta J/\psi$ as well as background from mis-identified electrons which originate from $\gamma$-conversion.
Due to the boost of the $\eta$ meson in the laboratory frame, the opening angles of its decay products are small.
The reaction $\EE\to\gamma\x$ recently observed by BESIII~\cite{bes3x3872}, where the photon comes from a radiative transition of the $Y(4260)$, represents an irreducible background to our signal process. To avoid this background, the ISR photon is required to be emitted at small polar angles $|\cos\theta_{\rm ISR}|>0.95$, almost colinear to the beam direction. Since the ISR photon cannot be detected in this region of the detector, its energy and polar angle are calculated from the missing momentum of the event (untagged ISR photon). As the photon from the radiative decay channel is predominantly emitted at large polar angles, an optimal signal to background ratio is obtained in this way. An MC simulation study shows that the $Y(4260)\to\gamma\x$ background can be neglected in the region of small polar angles of the ISR photon.
To improve the resolution of the $\pi^{+}\pi^{-}J/\psi$ mass spectrum and to further remove background, a two-constraint (2C) kinematic fit under the hypothesis of the $\gamma_{\rm ISR}\pi^{+}\pi^{-}\ell^{+}\ell^{-}$ final state is performed. The two constraints are the $J/\psi$ mass for the lepton pair and the mass of the missing ISR photon, which is zero. We accept events with $\chi^2_{2C}<15$.

\section{$\pi^+\pi^-J/\psi$ Mass Spectrum}
\begin{figure*}[htb]
\centering
\includegraphics[width=\linewidth]{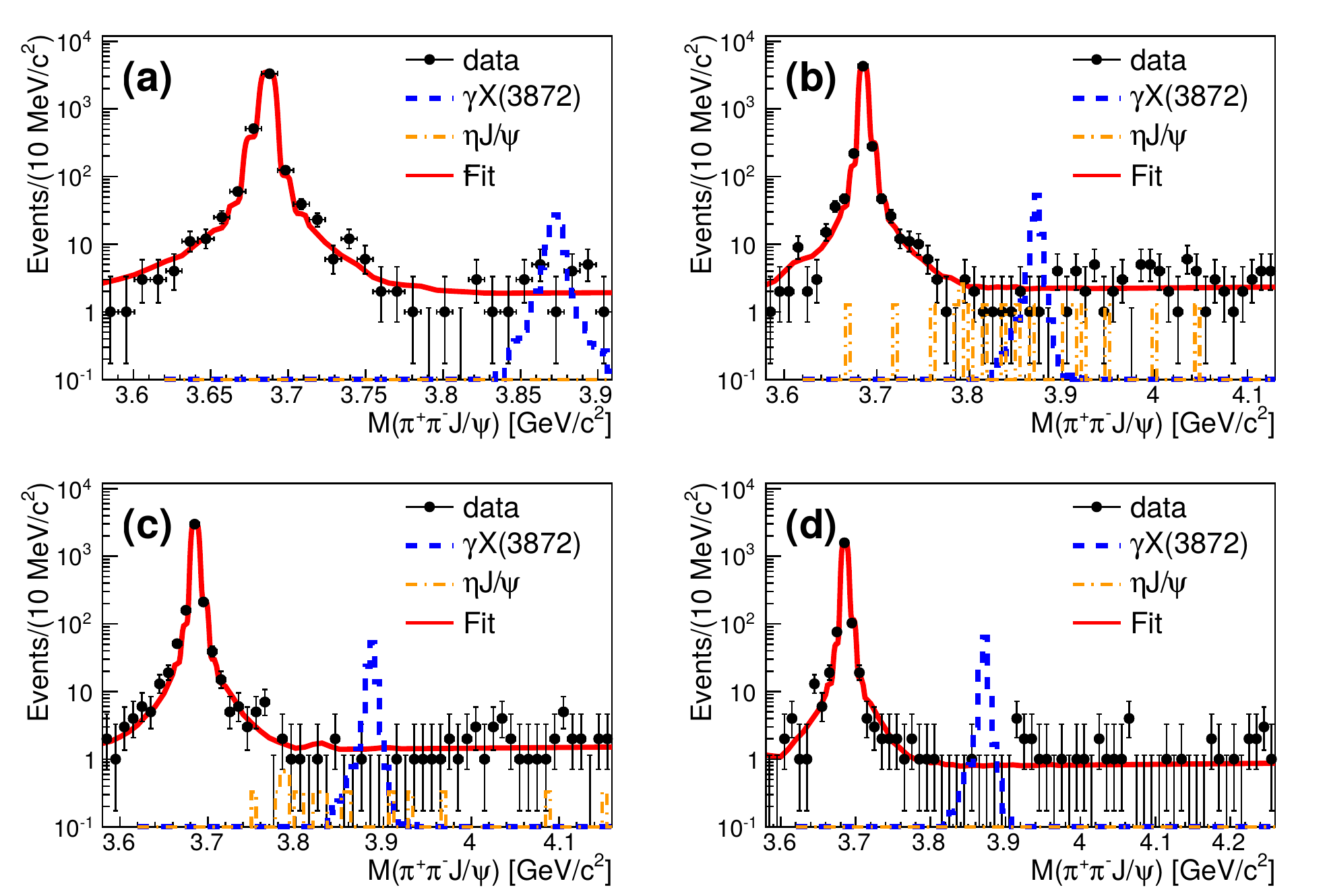}
\caption{The $\pi^+\pi^-J/\psi$ mass distributions at (a) $\sqrt{s}=4.009\,{\rm GeV}$, (b) $4.230\,{\rm GeV}$, (c) $4.260\,{\rm GeV}$ and (d) $4.360\,{\rm GeV}$. Dots with error bars
are data, the solid red lines are the fit curves, the blue dashed histograms are MC simulated $\x$ signal events, which are normalized arbitrarily, and the orange dot-dashed histograms are MC simulated $\eta J/\psi$ background events.}\label{massspectrum}
\end{figure*}

The invariant mass distributions of $M(\pi^+\pi^-J/\psi)$ for data, signal simulation, and simulation of the dominant background $\EE\to\eta J/\psi$ are shown in Fig.~\ref{massspectrum}.
All the selection criteria described above have been applied here.
As expected, the mass spectrum is dominated by the $\psi(3686)$ resonance.
No significant $\x$ peak is observed at any of the four c.m.\ energies. Hence, we set an upper limit for the electronic width of $\x$. In Fig.~\ref{massspectrum}, the blue dotted histogram represents the signal simulation of the $X(3872)$ with arbitrary normalization. The background channels of $\EE\to\pi^+\pi^-\pi^+\pi^-\gamma_{\rm ISR}$ and $\EE\to\eta^{\prime} J/\psi$ with $\eta^\prime\to\gamma\pi^+\pi^-$ are found to be negligible in an MC simulation study. The background channel $\EE\to\eta J/\psi$ with $\eta\rightarrow\pi^+\pi^-\pi^0$ is displayed as the orange dashed-dotted line in Fig.~\ref{massspectrum}.

Unbinned maximum likelihood fits are performed to extract the yields of $\psi(3686)$ and $\x$ events at each c.m.\ energy, where the line shapes of background are represented by polynomial functions and the line shapes of $\psi(3686)$ and $\x$ are described by the MC shape convoluted with a Gaussian function which takes into account resolution differences between data and MC simulation.
We use the same parameters of the Gaussian function for the two resonances. The fit results are displayed as the solid red curves in Fig.~\ref{massspectrum}. The event yields of $\psi(3686)$ from the fits are shown in Table~\ref{bwvalues}.

\section{Calculation of $\Gamma_{ee}$}
The measured radiative event yield $N_A$ of the process $\EE\to\gamma_{\rm ISR}A$ can be expressed as a function of $x \equiv 1-\frac{M(\pi^+\pi^-J/\psi)^2}{s}$~\cite{Druzhinin:2011qd}:
\begin{linenomath}
\begin{equation}\label{isrcrosssection}
\frac{dN_A}{dx}=W(s,x)\varepsilon_A\mathcal{L}\,\sigma(\EE\rightarrow A)\mathcal{B}(A\rightarrow f)\,,
\end{equation}
\end{linenomath}
where $s$ is the squared c.m.\ energy, $W(s,x)$ denotes the radiator function, $\varepsilon_A$ is the corresponding reconstruction efficiency, $\mathcal{L}$ is the integrated luminosity, $\sigma(\EE\rightarrow A)$ is the Born cross section to produce $A$ in $\EE$ annihilation, \mbox{$\mathcal{B}(A\rightarrow f)=\mathcal{B}(A\rightarrow \pi^+\pi^-J/\psi)\mathcal{B}(J/\psi\rightarrow\ell^+\ell^-)$} is the product of the branching fractions of $A$ decaying into the final state $f$.

The relationship between the electronic width $\Gamma_{ee}$ and the Born cross section reads:
\begin{linenomath}
\begin{equation}
\sigma(\EE\rightarrow A)=\frac{12\pi\Gamma_{ee}\Gamma_{\rm tot}}{(s^\prime-M_A^2)^2+\Gamma_{\rm tot}^2M_A^2}\,,
\end{equation}
\end{linenomath}
where $s^\prime=(1-x)s$, $\Gamma_{ee}$ ($\Gamma_{\rm tot}$) is the electronic (total) width of the resonance $A$, and $M_A$ is its mass. Eq.~(\ref{isrcrosssection}) must be integrated over $s^\prime$ in an appropriate region around the resonance $A$.
The integral only involves the Breit-Wigner function in the Born cross section and the radiator function. Hence it can be separated from the quantities determined in the measurement, such that the integral enters the calculation of the electronic width as a factor denoted by $I_A$. This factor is given by
\mbox{$I_A=12\pi\Gamma_{\rm tot}\int_{x_1}^{x_2}dx\frac{W(s,x)}{(s^\prime-M_A^2)^2+\Gamma_{\rm tot}^2M_A^2}$}.
The limits of the integral are chosen to coincide with the signal region.

Using Eq.~(\ref{isrcrosssection}), the electronic width times the branching fraction $\mathcal{B}(A\to\pi^+\pi^-J/\psi)$ can then be obtained via the relation
\begin{linenomath}
\begin{equation}\label{bwmethod}
\Gamma_{ee}^A\mathcal{B}(A\to\pi^+\pi^-J/\psi)=
\dfrac{N_A}{\varepsilon_A\mathcal{L}\,I_A\mathcal{B}(J/\psi\rightarrow\ell^+\ell^-)},
\end{equation}
\end{linenomath}
which is used to determine the electronic widths of $\x$ and $\psi(3686)$. As no significant signal is found in the case of $\x$, we calculate an upper limit for $\GEE^{\x}$. For the branching fractions we take the latest BESIII values $\mathcal{B}(\psi(3686)\rightarrow\pi^+\pi^-J/\psi)=(34.98\pm 0.45)\%$ and $\mathcal{B}(J/\psi\rightarrow\ell^+\ell^-)=(11.96\pm 0.05)\%$ ~\cite{Ablikim:2013pqa}.
The reconstruction efficiencies $\varepsilon_A$ are extracted from the signal MC sample $\EE\to\gamma_{\rm ISR}\x$ and $\EE\to\gamma_{\rm ISR}\psi(3686)$, respectively.
We apply an additional relative correction factor of 2\%, which stems from a data-MC difference found in the $\chi^2$ distributions. 
To obtain this correction factor, the number of events in the background-free $\psi(3686)$ mass region 
$(3.62<M(\pi^+\pi^-J/\psi)<3.75\,\,\text{GeV}/c^2)$  passing the $\chi^2_{2C}<15$ requirement relative to all reconstructed events in MC is compared to the respective number obtained from data.
All the values for the efficiencies and the integrals $I_A$ at each c.m.\ energy point are listed in Table~\ref{bwvalues}. The statistical errors of the efficiencies are negligible.
\begin{table*}[t]
\begin{center}
\caption{Values for the integrals ($I_{\psi(3686)}$ and $I_{\x}$), the efficiencies ($\epsilon_{\psi(3686)}$ and $\epsilon_{\x}$), the event yield $N^{obs}_{\psi(3686)}$ and the electronic widths ($\Gamma_{ee}^{\psi(3686)}$ and $\Gamma_{ee}^{X(3872)}\mathcal{B}(\x\to\pi^+\pi^-J/\psi)$). The errors shown are statistical only.}
\label{bwvalues}
\begin{tabular}{c|cccc}
\hline\hline
c.m.\ energy [GeV] & 4.009 & 4.230 & 4.260 & 4.360 \\
\hline
$\mathcal{L}\,[\text{pb}^{-1}]$         & $482$        & $1092$       & $826$        & $540$\\
$I_{\psi(3686)}\,[\text{pb/keV}]$ & $310$        & $172$        & $161$        & $133$\\
$I_{\x}\,[\text{pb/keV}]$         & $671$        & $247$        & $225$        & $174$\\
$\varepsilon_{\psi(3686)}$        & $0.303$      & $0.286$      & $0.286$      & $0.282$\\
$\varepsilon_{\x}$                & $0.314$      & $0.324$      & $0.325$      & $0.327$\\
$N^{\psi(2S)}$                    & $4168\pm 65$ & $5026\pm 71$ & $3547\pm 60$ & $1846\pm 43$\\
\hline
$\Gamma_{ee}^{\psi(3686)}\,[\text{eV}]$ & $2198 \pm 34$ & $2232 \pm 32$ & $2223 \pm 38 $ & $2176 \pm 51$\\
$\Gamma_{ee}^{X(3872)}\mathcal{B}(\x\to\pi^+\pi^-J/\psi)$ at 90\% C.L. [eV] & $0.630$  & $0.314$ & $0.319$ & $0.646$ \\
\hline\hline
\end{tabular} 
\end{center}
\end{table*}
First we compute the electronic width of $\psi(3686)$, which is denoted by $\Gamma_{ee}^{\psi(3686)}$. This serves as a benchmark and validation of our method, since the electronic width of $\psi(3686)$ is already known with high accuracy~\cite{PDG}. Applying the numbers for $\psi(3686)$ listed in Table~\ref{bwvalues} to Eq.~(\ref{bwmethod}), we obtain the value for $\Gamma_{ee}^{\psi(3686)}$ at each of the four energy points separately, as shown in Table~\ref{bwvalues}.
We calculate the error weighted average of the electronic width of $\psi(3686)$ from the four single measurements in Table~\ref{bwvalues}, which gives 
$\Gamma_{ee}^{\psi(3686)}=\left(\geepsip\pm \statErrpsip_\text{stat}\right)\,\text{eV}\,.$

Since no $\x$ signal is observed, we set an upper limit at the 90\% confidence level (C.L.) for its electronic width. Applying the Bayesian method, we perform likelihood scans at each of the four data sets of the electronic width times the branching fraction, which is proportional to the $X(3872)$ event yield parameter $N_i$ according to Eq.~(\ref{bwmethod}). This provides four likelihood curves, that are denoted by $L_i(\gamma)\,,\,i=1\ldots 4$, where \mbox{$\gamma=\Gamma_{ee}^{X(3872)}\mathcal{B}(\x\to\pi^+\pi^-J/\psi)$}.
We look for the values $\gamma_{i}^{\rm up}$ that yield 90\% of the likelihood integral over $\gamma$ from zero to infinity: \mbox{$\int_0^{\gamma_{i}^{\rm up}}d\gamma L_i(\gamma)=0.9\int_0^{\infty}d\gamma L_i(\gamma)$}. 
In order to combine the four measurements, we construct the likelihood of the combined measurement.
The four single likelihood curves are scaled such that they have the same value at their respective maxima.
We take the product of the likelihood scan curves of the single measurements.
The upper limit $\gamma^{\rm up}_{\rm tot}$ at the 90\% C.L. of $\gamma$ is determined from
\begin{linenomath}
\begin{equation}
\int_0^{\gamma^{\rm up}_{tot}}d\gamma
\prod_{i=1}^{4}L_i(\gamma)=0.9\int_0^{\infty}d\gamma
\prod_{i=1}^{4}L_i(\gamma)\,,\nonumber
\end{equation}
\end{linenomath}
We obtain 
\mbox{$\gamma^{\rm up}_{tot}=\Gamma_{ee}^{X(3872)}\mathcal{B}(\x\to\pi^+\pi^-J/\psi)$} $=\geBrXnoSys\,\text{eV}$ at the 90\% C.L. 

\section{Estimation of Systematic Uncertainties}\label{sec:error}
The luminosity is measured using large angle Bhabha events, and the uncertainty is estimated to be 1\%~\cite{lumi}. The uncertainty related to the tracking efficiency is 1\% per charged track~\cite{bes3x3872}. Since the final state has four charged tracks, we estimate an uncertainty of 4\% for the whole event.
Applying our $J/\psi$ selection both to data and the $\psi(3686)\gamma_{\rm ISR}$ MC simulation, the obtained event yield differs by 0.2\%, which we take as systematic uncertainty for the $J/\psi$ selection.
To correct for differences between data and MC simulation in the $\chi^2_{2C}$ distribution, an efficiency correction was determined. Varying the $\chi^2_{2C}$ selection and calculating the efficiency correction factor again at each energy, we obtain a corresponding uncertainty of 0.4\% in the luminosity weighted average.
The integrals $I_A$ have an uncertainty of 0.7\%, due to the precision of the numerical integration (0.5\%)
and the calculation of the radiator function (0.5\%).
The relative uncertainties of the branching fraction $\mathcal{B}(\psi(3686)\rightarrow\pi^+\pi^-J/\psi)$ and $\mathcal{B}(J/\psi\rightarrow\ell^+\ell^-)$ are 1.3\% and 0.5\%, respectively. There is no correlation between these branching fractions~\cite{Ablikim:2013pqa}. We take 1.4\% as the systematic uncertainty from the branching fractions for the electronic width of $\psi(3686)$. In the calculation of $\Gamma_{ee}^{X(3872)}\mathcal{B}(\x\to\pi^+\pi^-J/\psi)$ only the branching fraction 
$\mathcal{B}(J/\psi\rightarrow\ell^+\ell^-)$ appears. Hence, the corresponding uncertainty is 0.5\%~.
To estimate the systematic uncertainty due to the width assumed for $\x$, we change the width by $\pm 0.2\,\text{MeV}/c^2$ and repeat the entire fitting procedure.
The maximal relative difference of these results from the result obtained with the standard width is found to be 2.7\% in the luminosity-weighted average. 
The detection efficiency of ISR $\x$ events was determined from a MC simulation using the {\sc vectorisr} model~\cite{VECISR}, since this final state is not available in the {\sc phokhara} event generator.
On the other hand, the ISR $\psi(3686)$ detection efficiency was determined using the {\sc phokhara} event generator, which simulates ISR events with 0.5\% precision. 
To obtain the uncertainty of the ISR simulation with the {\sc vectorisr} model, we compare the efficiencies of 
ISR $\psi(3686)$ events generated with the {\sc phokhara} event generator~\cite{Czyz:2010hj} and 
the {\sc vectorisr} module~\cite{VECISR}. The luminosity-weighted average difference is found to be 
3.4\% between them, which is taken as systematic uncertainty for the {\sc vectorisr} model.
\vspace{1mm}
\begin{Figure}
\centering
\captionof{table}{Sources of systematic uncertainties and their contribution (\%).}
\label{sysErrSummary}
\begin{tabular}{c|cc}
\hline\hline
Source & $\sigma_{\rm sys}^{X(3872)}$ & $\sigma_{\rm sys}^{\psi(3686)}$\\ 
\hline 
Luminosity           & $1.0$   & $1.0$\\  
Tracking             & $4.0$   & $4.0$\\ 
$J/\psi$ selection   & $0.2$   & $0.2$\\  
Kinematic Fit        & $0.4$   & $0.4$\\  
Integrals $I_A$      & $0.7$   & $0.7$\\
Branching ratio      & $0.5$   & $1.4$\\ 
$\x$ width           & $2.7$   & -\\
ISR simulation       & $3.4$   & -\\ 
$\psi(3686)$ fit model &  -    & $1.0$ \\
\hline
Total &  \sysx & \syspsip \\
\hline\hline
\end{tabular} 
\end{Figure}
For $\Gamma_{ee}^{\psi(3686)}$ a further systematic uncertainty occurs due to the choice of the fit function. In order to deal with this uncertainty, we determine the number of $N^{\text{MC}}_{\psi(3686)}$ using a second fit function, which is a double Gaussian for the $\psi(3686)$ peak plus a Gaussian for the $\x$ plus a constant  for background.
In the luminosity-weighted average, this fit model differs by 1.0\%, which is taken as systematic uncertainty.
Signal events with a hard final state radiation (FSR) photon are rejected since the $J/\psi$ mass is constraint in the kinematic fit. Thus FSR effects are negligible.
Systematic uncertainties from the background shape and the fit range have been found to be negligible.
The full list of systematic uncertainties is shown in Table~\ref{sysErrSummary}.
Assuming the sources to be independent, the total systematic uncertainty for the electronic width of $\x$ is $\sysx\%$,
while in the case of $\psi(3686)$ we find a systematic uncertainty of $\syspsip\%$.

\section{Summary}
We have performed a search of the process $\EE\to\gamma_{\rm ISR}\x\to\gamma_{\rm ISR}\pi^{+}\pi^{-}J/\psi$ using the ISR untagged method, where the production of $\x$ in $\EE$ annihilations is possible via a two-photon box diagram. No significant $\x$ signal is observed in the $\pi^{+}\pi^{-}J/\psi$ mass spectrum. We set an upper limit for the electronic width of $\x$. By combining all four data sets, we finally obtain
\begin{linenomath}
\begin{equation}
\Gamma_{ee}^{X(3872)}\mathcal{B}(\x\to\pi^+\pi^-J/\psi)<\geBrX\,\text{eV}\nonumber
\end{equation}
\end{linenomath}
at the 90\% C.L. Here we have multiplied the upper limit by a factor $1/(1-\sigma_{sys})$ in order to take the systematic uncertainties into account.
Our measurement improves upon the current limit
\mbox{$\Gamma_{ee}^{X(3872)}\mathcal{B}(\x\to\pi^+\pi^-J/\psi)<6.2\,\text{eV}$} at the 90\% C.L.~\cite{Aubert:2005eg} by a factor of \improve. If we assume the branching fraction \mbox{$\mathcal{B}(X(3872)\rightarrow\pi^+\pi^-J/\psi)>3\%$} \cite{PDG,Yuan:2009iu}, we obtain an upper limit for the electronic width of $\x$ to be \mbox{$\Gamma_{ee}^{X(3872)}<\geX\,\text{eV}$}.
For the first time we obtain a value for $\GEE^{\x}$ on the $\mathcal{O}(\text{eV})$~level, which is the level predicted for ordinary charmonium states~\cite{Kuhn:1979bb}.
However, our upper limit is still larger than a theoretical calculation~\cite{Denig:2014fha} which predicts $\GEE\gtrsim 0.03~\text{eV}$. The results should encourage theorists to compute the electronic width of $\x$ under different assumptions regarding its intrinsic nature and to confront these calculations with our measurement. This might lead to new insights regarding the nature of $\x$.

We have also measured the electronic width of the well-known $\psi(3686)$ resonance with the result: 
\begin{linenomath}
\begin{equation}
\Gamma_{ee}^{\psi(3686)}=\left(\geepsip\pm\statErrpsip_\text{stat}\pm\sysErrpsip_\text{sys}\right)\,\text{eV}\,.\nonumber
\end{equation}
\end{linenomath}
This is in agreement with the PDG~\cite{PDG} fit, which is $\left(2360\pm 40\right)\,\text{eV}$. 
With a similar accuracy as the one reported in~\cite{Ablikim:2008zzb}, this is the
best individual measurement of $\GEE^{\psi(3686)}$ to date.

\section{Acknowledgement}
The BESIII collaboration thanks the staff of BEPCII and the IHEP computing center for their strong support. This work is supported in part by National Key Basic Research Program of China under Contract No.~2015CB856700; National Natural Science Foundation of China (NSFC) under Contracts Nos.~11125525, 11235011, 11322544, 11335008, 11425524; the Chinese Academy of Sciences (CAS) Large-Scale Scientific Facility Program; Joint Large-Scale Scientific Facility Funds of the NSFC and CAS under Contracts Nos.~11179007, U1232201, U1332201; CAS under Contracts Nos.~KJCX2-YW-N29, KJCX2-YW-N45; 100 Talents Program of CAS; INPAC and Shanghai Key Laboratory for Particle Physics and Cosmology; German Research Foundation DFG under Contract No.~CRC-1044;
Seventh Framework Programme of the European Union under Marie Curie International Incoming Fellowship Grant Agreement No.~627240; Istituto Nazionale di Fisica Nucleare, Italy; Ministry of Development of Turkey under Contract No.~DPT2006K-120470; Russian Foundation for Basic Research under Contract No.~14-07-91152; U. S. Department of Energy under Contracts Nos.~DE-FG02-04ER41291, DE-FG02-05ER41374, DE-FG02-94ER40823, DESC0010118; U.S. National Science Foundation; University of Groningen (RuG) and the Helmholtzzentrum fuer Schwerionenforschung GmbH (GSI), Darmstadt; WCU Program of National Research Foundation of Korea under Contract No.~R32-2008-000-10155-0.

\end{multicols}

\end{document}